\documentclass[preprint,showpacs,12pt]{revtex4}
\usepackage{amsmath,amssymb,amsfonts}
\usepackage{graphicx}
\usepackage{hhline}
\usepackage{bm}
\usepackage{amsmath}
\usepackage{epsfig}

\begin{document}
\begin{flushleft}
CERN-PH-TH/2013-060\\
\end{flushleft}
\title{Quasielastic and multinucleon excitations in antineutrino-nucleus interactions}
\author {M. Martini}
\affiliation{Institut d'Astronomie et d'Astrophysique, CP-226, Universit\'e Libre de Bruxelles, 1050 Brussels, Belgium}
\author {M. Ericson}
\affiliation{Universit\'e de Lyon, Univ. Lyon 1,
 CNRS/IN2P3, IPN Lyon, F-69622 Villeurbanne Cedex, France}
\affiliation{Physics Department, Theory Unit, CERN, CH-1211 Geneva, Switzerland}

\begin{abstract}
 We investigate the MiniBooNE recent data on the antineutrino nucleus interaction, using the same theoretical description with the same parameters as in previous works on neutrino interactions. The
double differential  quasielastic cross section, which is free from the energy reconstruction problem,  is well reproduced by our model once the multinucleon excitations are incorporated. A similar agreement is achieved for the $Q^2$ distribution.  
\end{abstract}

\maketitle

\section{Introduction}
The recent publication by the MiniBooNE group of the antineutrino charged-current (CC) quasielastic cross section on $^{12}$C \cite{AguilarArevalo:2013hm} 
completes the neutrino data \cite{AguilarArevalo:2010zc,AguilarArevalo:2010cx} 
allowing  a full confrontation of the theoretical descriptions with the experimental data. 
For neutrinos a successful description of the quasielastic cross section needs the inclusion of the multinucleon component which, 
in a Cerenkov detector is indistinguishable from the genuine quasielastic part \cite{Martini:2009uj}. 
Its introduction allows a successful reproduction of the data without any modification of the nuclear 
axial form factor. 
The aim of the present work is to test our theoretical description in the different situation provided by the antineutrino interaction, keeping on purpose 
exactly the same parameters of previous works which successfully reproduce the experimental data \cite{Martini:2009uj,Martini:2010ex,Martini:2011wp}. 
The most significant one is the double differential cross section \cite{Martini:2011wp} function of 
two measured quantities, the muon energy and the scattering angle, hence free from the energy reconstruction problem 
\cite{Martini:2012fa,Martini:2012uc,Lalakulich:2012ac,Lalakulich:2012hs,Nieves:2012yz}.  
We briefly summarize the essence of our model which is described in details in \cite{Martini:2009uj} and in \cite{Martini:2010ex} for antineutrinos. 
Our description treats the genuine quasielastic cross section in the random phase approximation (RPA) scheme. 
For the multinucleon part our treatment is based on the work by Alberico \textit{et al.} \cite{Alberico:1983zg} which aims at the description of 
the $(e,e')$ transverse response and in particular the filling of the dip between the quasielastic and Delta excitations. 
Alberico \textit{et al.} \cite{Alberico:1983zg} 
interpreted this filling as originating from the two particle-two hole excitations of the nuclear system by the virtual photon. 
As for the part which represents the non pionic in medium decay of the Delta, it is taken from the parameterization of Oset \textit{et al.} \cite{Oset:1987re}. 
The work of Alberico \textit{et al.} concerned exclusively the magnetic response, which, by virtue of the couplings, is  of isovector nature.
For our work on neutrinos, the important observation is that the longitudinal, or charge response, in $(e,e')$ scattering instead 
does not display an evidence for a cross section excess above the quasielastic peak. 
This is confirmed by the superscaling analysis of electron scattering data \cite{Maieron:2001it,Martini:2007jw}. 
The various components which build the neutrino cross sections are excited by the isovector component of the charge operator, 
or by the nucleon spin -isospin operators (see Eq. (1) of \cite{Martini:2010ex}). 
Motivated by these observations we have introduced the two particle-two hole excitations exclusively in the spin-isospin channels, which is 
a distinct feature of our description. 
Due to the axial-vector interference term the spin-isopin contribution weights less for antineutrinos.
The consequence is that the multinucleon piece should weight less on the cross section for antineutrinos than for neutrinos. 
This is not the case in other approaches \cite{Amaro:2010sd,Amaro:2011aa,Nieves:2011pp,Nieves:2013fr,Meucci:2011vd,Meucci:2012yq}. 
The model closest in spirit to our treatment is the one of Bodek \textit{et al.} \cite{Bodek:2011ps} characterized by a modification
of the magnetic form factor so as to account for the observed excess in the dip region of the magnetic response.
For a comparison between theoretical approaches see for example \cite{Martini:2011ui}.

\section{Analysis of the cross sections}

\begin{figure}
\begin{center}
  \includegraphics[width=16cm,height=12cm]{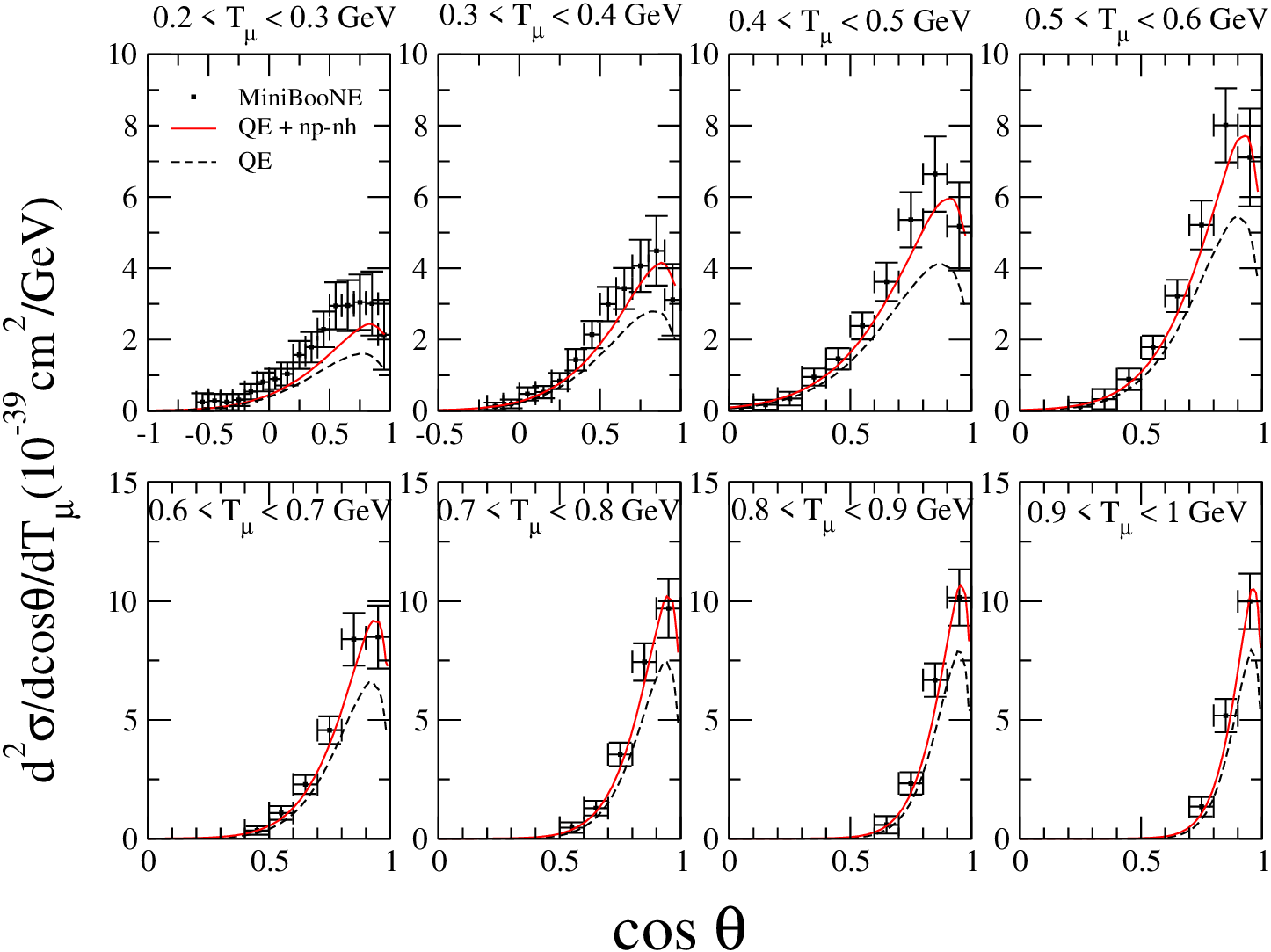}
\caption{(color online). MiniBooNE flux-averaged CC ``quasielastic'' $\bar{\nu}_\mu$-$^{12}$C double differential 
cross section per proton for several values of muon kinetic energy
as a function of the scattering angle. Dashed curve: pure quasielastic (1p-1h) cross section calculated in RPA; 
solid curve: with the inclusion of the multinucleon (np-nh) component. 
The experimental MiniBooNE points with the shape uncertainty are taken from \cite{AguilarArevalo:2013hm}. 
For the data there is an additional normalization uncertainty of 17.2\% not shown here.}
\label{fig_minib_d2s_anti}
\end{center}
\end{figure}

We first remind the expression of the double differential cross section which applies for neutrino as well as for antineutrino. 
For a given ``quasielastic'' event the muon energy $E_\mu$ (or kinetic energy $T_\mu$) and its emission angle
$\theta$ are measured while the neutrino energy $E_\nu$ is unknown. The expression of the double differential cross section in terms of the measured quantity is 

 \begin{equation}
 \label{cross}
\frac{d^2 \sigma}{dT_{\mu}~d\mathrm{cos}\theta}=
\frac{1}
{ \int \Phi(E_{\nu})~d E_{\nu}}
 \int ~d E_{\nu}
\left[\frac{d^2 \sigma}{d \omega  ~d\mathrm{cos}\theta}\right]_{\omega=E_{\nu}-E_{\mu}} \Phi(E_{\nu}).
\end{equation}
In the numerical evaluations we use the antineutrino flux $\Phi(E_{\nu})$ from Ref.\cite{AguilarArevalo:2013hm}. 
As in our work \cite{Martini:2011wp} we have applied  relativistic corrections to the nuclear responses. 

\begin{figure}
\begin{center}
  \includegraphics[width=12cm,height=8cm]{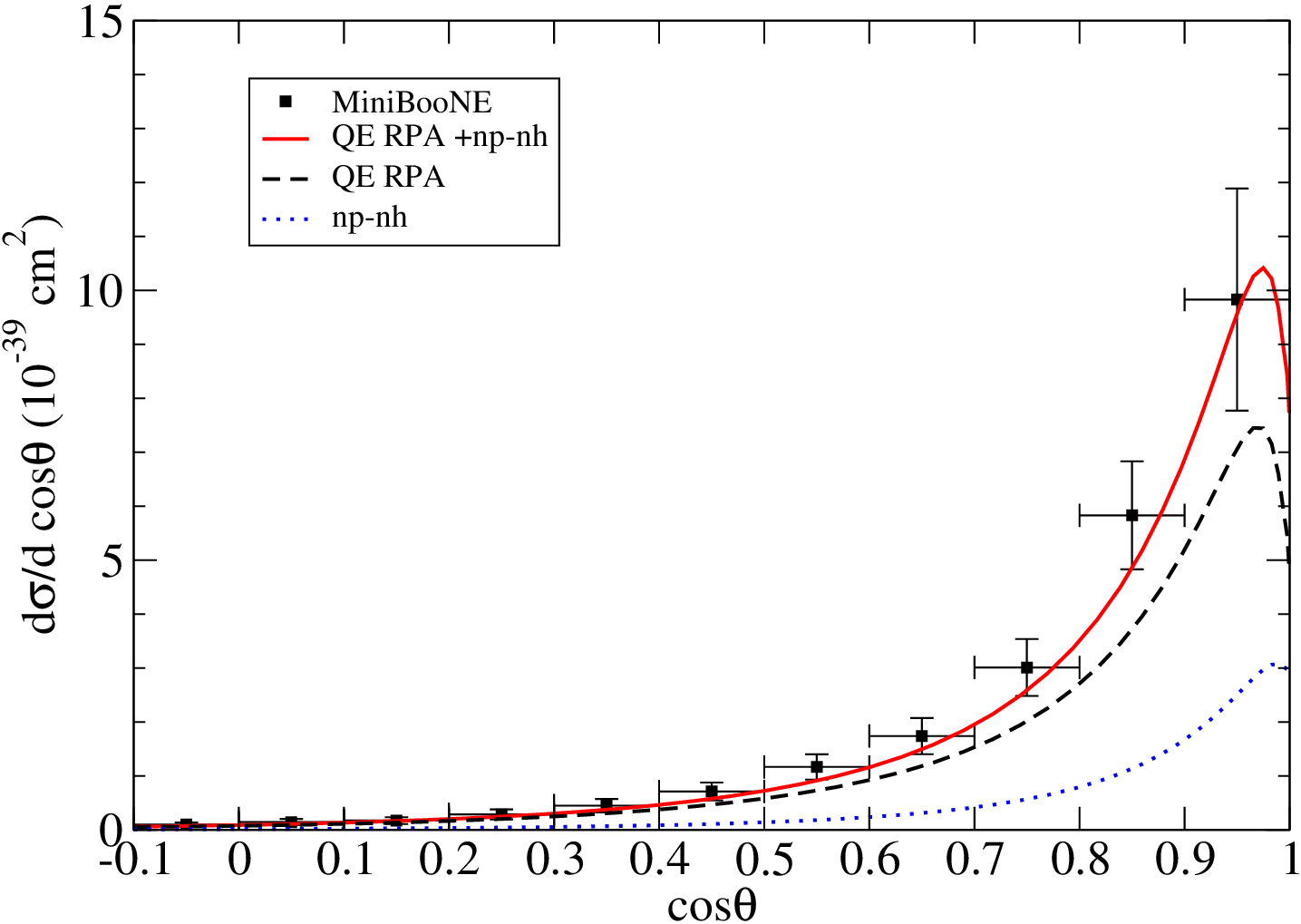}
\caption{(color online). MiniBooNE flux-averaged CC ``quasielastic'' $\bar{\nu}_\mu$-$^{12}$C differential cross section per proton 
as a function of the muon scattering angle. 
Note that in order to compare with data the integration is performed over the muon kinetic energies 0.2 GeV $<T_\mu<$  2.0 GeV.
Dashed curve: pure quasielastic (1p-1h) cross section; 
solid curve: with the inclusion of np-nh component; dotted line: np-nh contribution. 
The experimental MiniBooNE points with the shape uncertainty 
are taken from \cite{AguilarArevalo:2013hm}. There is an additional normalization uncertainty of 17.2\% not shown here.}
\label{fig_minib_ds_dcos_anti}
\end{center}
\end{figure}

\begin{figure}
\begin{center}
  \includegraphics[width=12cm,height=8cm]{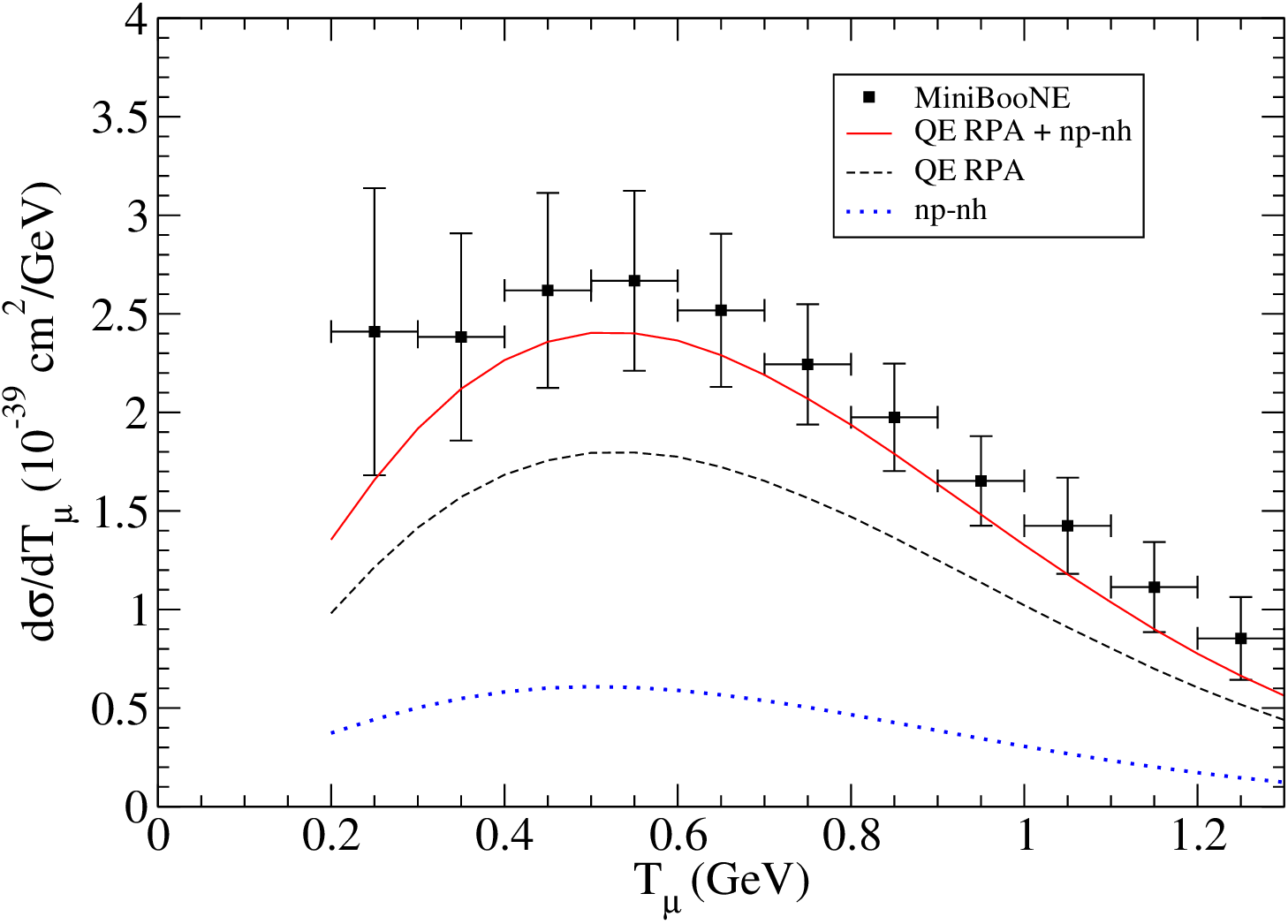}
\caption{(color online). MiniBooNE flux-averaged CC ``quasielastic'' $\bar{\nu}_\mu$-$^{12}$C differential cross section per proton 
as a function of the muon kinetic energy. Dashed curve: pure quasielastic (1p-1h) cross section; 
solid curve: with the inclusion of np-nh component; dotted line: np-nh contribution. 
The experimental MiniBooNE points with the shape uncertainty 
are taken from \cite{AguilarArevalo:2013hm}. There is an additional normalization uncertainty of 17.2\% not shown here.}
\label{fig_minib_ds_dt_anti}
\end{center}
\end{figure}

\begin{figure}
\begin{center}
  \includegraphics[width=12cm,height=8cm]{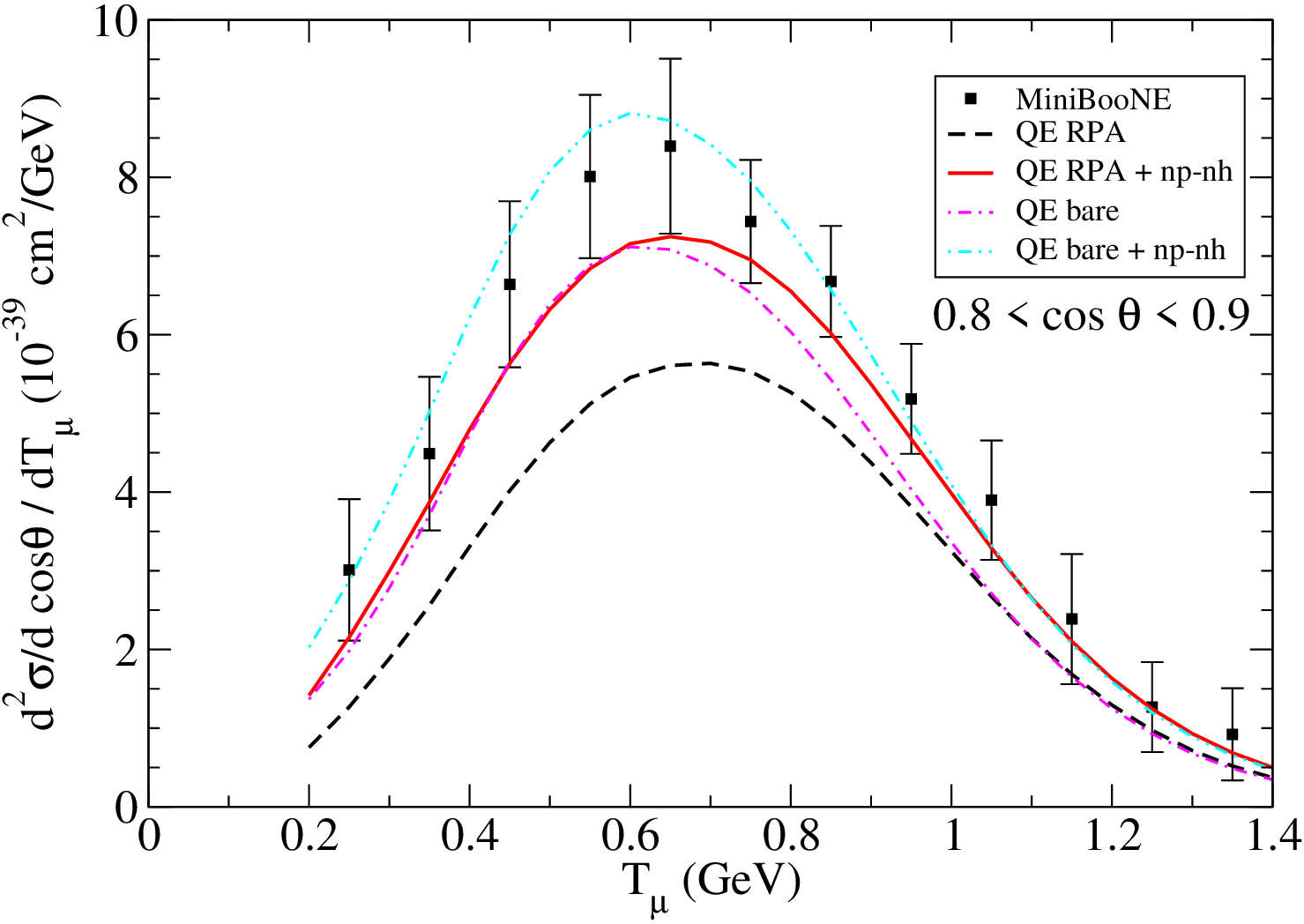}
\caption{(color online). MiniBooNE flux-averaged CC quasielastic $\bar{\nu}_\mu$-$^{12}$C 
double differential cross section per proton for 0.8 $<$ cos$\theta~<$ 0.9
as a function of the muon kinetic energy. 
Dashed curve: pure quasielastic calculated in RPA; solid curve: RPA quasielastic 
with the inclusion of np-nh component; dot-dot-dashed: bare quasielastic with the inclusion of np-nh component; 
dot-dashed curve: bare quasielastic. The experimental MiniBooNE points with the shape uncertainty are taken from \cite{AguilarArevalo:2013hm}. 
There is an additional normalization uncertainty of 17.2\% not shown here.} 
\label{fig_minib_d2s_cos_fix_vs_tmu_anti}
\end{center}
\end{figure}

The results of the double differential cross section are displayed 
in Fig. \ref{fig_minib_d2s_anti}, 
with and without the inclusion of the multinucleon (np-nh) component and compared to the experimental data \cite{AguilarArevalo:2013hm}. 
A similar comparison have been recently reported in \cite{Nieves:2013fr}. 
Our evaluation, as all those of this article, is done with the free value of the axial mass. 
The agreement between our predictions and the data is quite good in all the measured range once the multinucleon component is incorporated,  
which is remarkable in view of the fact that no parameter has been changed with respect to our calculations in the neutrino mode.   
The only panel presenting some disagreement, of which we do not know the origin, corresponds to the lowest $T_\mu$ values, 0.2 MeV$<T_\mu<$ 0.3 MeV, 
where the theoretical prediction is lower than the experimental data. 
Notice that this underestimation at low $T_\mu$ has little influence on the once integrated quantity $d\sigma/d\cos\theta$ 
shown in Fig. \ref{fig_minib_ds_dcos_anti}, while 
Fig. \ref{fig_minib_ds_dt_anti} displays the quantity $d\sigma/dT_\mu$.  
In both cases we are fully compatible with the experimental results. Nevertheless a small but systematic underestimation shows up with respect to data, 
at least in the present normalization. We remind the additional normalization uncertainty of 17.2\% in data \cite{AguilarArevalo:2013hm}. 
Within this error margin we are in excellent agreement.
We observe in Fig. \ref{fig_minib_ds_dcos_anti} that the antineutrino cross section falls more rapidly with angle than the neutrino one 
(compare with Fig. 9 of \cite{Martini:2011wp}). This also reflects in the $Q^2$ distribution which peaks at smaller  $Q^2$ values than the neutrino one. 
The double differential cross sections as a function of $T_\mu$ for 0.8 $<$cos$\theta<$0.9 is displayed in Fig. \ref{fig_minib_d2s_cos_fix_vs_tmu_anti}. 
It manifests the same systematical underevaluation trend.
We have chosen this angle band to be able to compare with the similar curve for neutrinos (Fig. 6 of \cite{Martini:2011wp}). 
It happens that for this band  the theoretical underevalution is the most pronounced 
(see the corresponding point in Fig. \ref{fig_minib_ds_dcos_anti}). 
As this trend is nevertheless present we may investigate its origin. 
On a purely theoretical round, we describe the genuine quasielastic cross section in RPA where the repulsive particle-hole interaction has a quenching effect 
\cite{Alberico:1981sz}. 
In Fig. \ref{fig_minib_d2s_cos_fix_vs_tmu_anti} this RPA quenching explicitely appears by comparing the cross sections with and without RPA. 
We remind that for neutrinos the RPA effect is needed in order to reproduce the double differential cross sections as well as the $Q^2$ distribution
\cite{Martini:2011wp}. The only freedom that we have for antineutrinos is then on the RPA effect of the isovector response. It does not affect the neutrino cross sections in view of the small weight of this response.
We have then investigated the influence of this RPA suppression in  the isovector response. 
It has no effect for neutrinos and even for antineutrinos it is also too small to produce a significant increase of the cross section. It does not offer an issue for the slight but systematic theoretical underevaluation trend. It seems that this has to be found rather in the data uncertainty which  is 17.2\%. A reduction of the data by this amount is sufficient to make the agreement theory-experiment excellent, as good as for neutrinos.

\begin{figure}
\begin{center}
  \includegraphics[width=12cm,height=8cm]{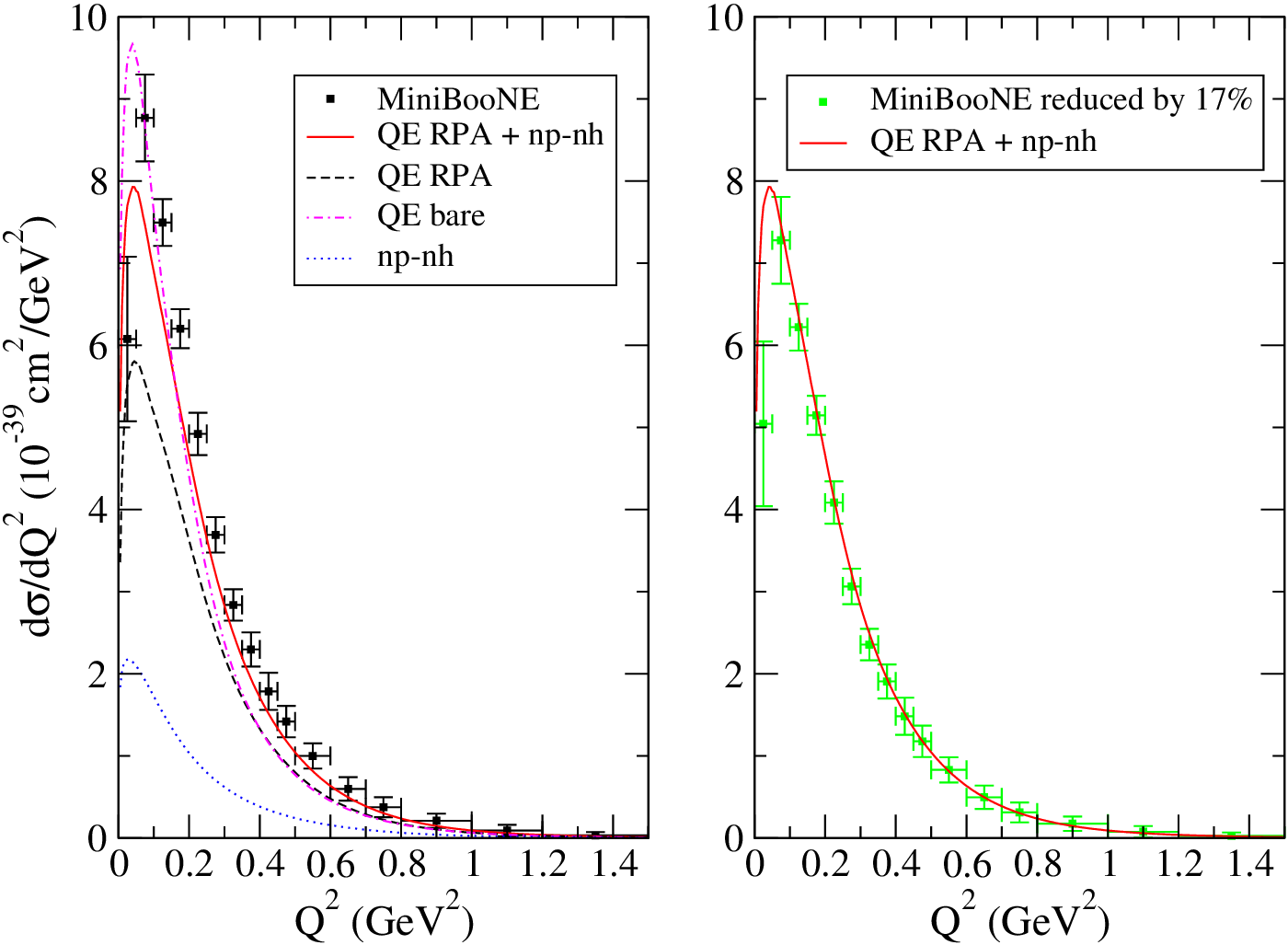}
\caption{(color online). MiniBooNE flux-averaged $\bar{\nu}_\mu$ CC $Q^2$ distribution per proton. Dashed curve: pure quasielastic (1p-1h); 
solid curve: with the inclusion of np-nh component; dotted line: np-nh component; dot-dashed line: bare distribution. 
The experimental MiniBooNE points with the shape uncertainty
are taken from \cite{AguilarArevalo:2013hm}. For the data there is an additional normalization uncertainty of 17.2\%. 
In the right panel a reduction of 17\% of the MiniBooNE data is performed.}
\label{fig_minib_cc_ds_dQ2_anti}
\end{center}
\end{figure}

The $Q^2$ distribution is shown in Fig. \ref{fig_minib_cc_ds_dQ2_anti} with and without the multinucleon component.  
The bare genuine quasielastic result is also shown. 
As for neutrino the RPA effects disappear beyond $Q^2\gtrsim 0.3$ GeV$^2$ where the presence of the multinucleon component is required. 
The agreement theory experiment is quite good. The experimental points are given in terms of the reconstructed value of  $Q^2$ while in our theory it is the real value. The influence of this difference has been shown to be small by Lalakulich \textit{et al.} \cite{Lalakulich:2012hs}. 
For information we show in the right panel of Fig. \ref{fig_minib_cc_ds_dQ2_anti} the effect on this distribution of a systematical reduction of the data by $17\%$. In this case the agreement becomes excellent,  as the one that we had for neutrinos.

\begin{figure}
\begin{center}
\includegraphics[width=12cm,height=8cm]{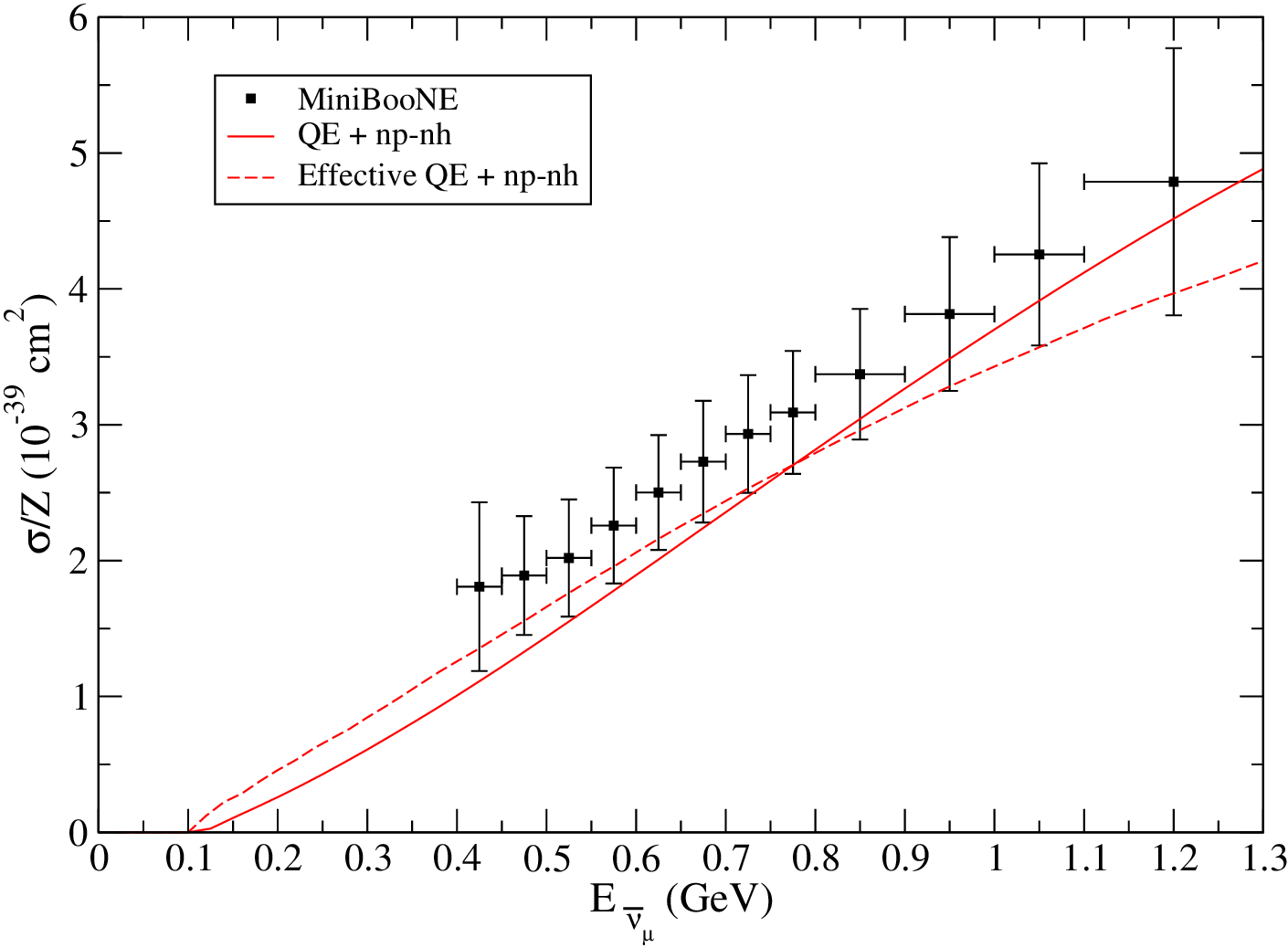}
\caption{(Color online) Theoretical (solid line) and effective (dashed line) $\bar{\nu}_{\mu}$-$^{12}$C cross section per proton including the multinucleon component. 
The experimental MiniBooNE result with the total error \cite{AguilarArevalo:2013hm} is also shown.
}
\label{fig_effective_sigma_muon_anti}
\end{center}
\end{figure}

Finally we discuss the case of the total cross section as a function of the antineutrino energy. 
We show it in Fig. \ref{fig_effective_sigma_muon_anti} together with experimental data. 
We remind that this experimental quantity is not model independent, contrary to the double differential cross section. 
Data are given as a function of the reconstructed neutrino energy and not of the genuine one. 
Hence one deals with an effective cross section which depends on the shape of the (anti)neutrino energy distribution. 
We have discussed in details the problem of the energy reconstruction in two recent works \cite{Martini:2012fa,Martini:2012uc}. 
Figure \ref{fig_effective_sigma_muon_anti} shows the influence of the energy reconstruction by comparing the effective cross section with 
the theoretical one, function of the true neutrino energy. The experimental data are also displayed. 
As in \cite{Martini:2012uc}, reconstruction produces some increase at low energy and lowers the cross section at large ones. 
We remind that this difference depends on the shape of the flux. Contrary to previous cases, 
here the error bar on the experimental points includes the renormalization uncertainty. Our theoretical curve is within the error band but on the low side, 
as expected from the trend of the various differential cross sections.

\section{Conclusion}
In this work we have investigated in detail the antineutrino - $^{12}$C cross sections in connection with MiniBooNE data. Our theoretical approach is, in all the aspects, identical to the one used in our previous works on neutrinos.
The most significant quantity is the double differential cross section which does not imply any reconstruction of the antineutrino energy. 
  For this quantity the agreement of our RPA approach with data is good once the np-nh component is included. 
We have also examined the $Q^2$ distribution which establishes the necessity of the multinucleon contribution, independently of the RPA quenching. It confirms our first suggestion that there is no need for a change in the axial mass once the multinucleon processes are taken into consideration. 
In spite of the identity of the inputs, which are the nuclear response functions, for neutrino and antineutrino calculations , 
we remind that the various responses weight differently in the respective cross sections, generating an asymmetry of the nuclear effects for neutrinos and antineutrinos. 
This is discussed in details in \cite{Martini:2010ex}. We suggested that the antineutrino cross section would offer a crucial test of our nuclear model. 
The conclusion of the presence investigation is that, after its success in the neutrino case, 
our model stands quite well the test of the comparison with the recent antineutrino data which are well reproduced by our theoretical description. 
With  a  $17\%$, reduction of the data, compatible with the given normalization uncertainty, an even better agreement, of the same quality as for neutrinos, 
could be reached.
  The asymmetry between neutrinos and antineutrinos interactions is important  for CP violation effects. We have shown that nuclear effects generate an additional asymmetry. It has been the object of the present work to test, with success, our understanding of this asymmetry. 

{\bf{Acknowledgments}}
\\
We thank Joe Grange for useful discussions. 
This work was partially supported by the Communaut\'e Fran\c caise de Belgique (Actions de Recherche Concert\'ees).  

\end{document}